\documentclass[12pt]{article}
\usepackage{graphicx}
\usepackage{epsfig}

\begin{document}

\title{{\bf Form factors in relativistic quantum mechanic: 
constraints from space-time translations}} 
\author{ 
B.  Desplanques$^{1}$\thanks{{\it E-mail address:}  desplanq@lpsc.in2p3.fr},
Y.B. Dong$^{2}$\thanks{{\it E-mail address:}  dongyb@mail.ihep.ac.cn}
\\
$^{1}$LPSC, Universit\'e Joseph Fourier Grenoble 1, CNRS/IN2P3, INPG,\\
%Institut National Polytechnique de Grenoble, \\ 
  F-38026 Grenoble Cedex, France \\
$^{2}$Institute of High Energy Physics, Chinese Academy of Science,\\ 
Beijing 100049, P. R. China}

\sloppy

\maketitle

\begin{abstract}
The comparison of form factors calculated from a single-particle 
current in different relativistic quantum mechanic approaches  evidences 
tremendous discrepancies. The role of constraints co\-ming from space-time 
translations is considered here with this respect. It is known 
that invariance under these translations implies the energy-momentum 
conservation relation that is usually assumed to hold globally.
Transformations of the current under these translations, which lead to this
result, also imply constraints that have been  ignored so far 
in relativistic quantum mechanic approaches. 
An implementation of these constraints is discussed in the case 
of a model with two scalar constituents. It amounts to incorporate 
selected two-body currents to all orders in the interaction. 
Discrepancies for form factors in different approaches can thus be removed, 
contributing to restore the equivalence of different approaches. 
Results for the standard front-form approach ($q^+=0$) are found 
to fulfill the constraints and are therefore unchanged. 
The relation with results from a dispersion-relation approach is also made.
\end{abstract}

%%%%%%%%%%%%%%%%%%%%%%%%%%%%%%%%%%%%%%%%%%%%%%%%%%%%%%%%%%%%%%%%%%%%%%%%%%%%%%%%%%%%%%%%%
%%%%%%%%%%%%%%%%%%%%%%%%%%%-11111111-%%%%%%%%%%%%%%%%%%%%%%%%%%%%%%%%%%%%%%%%%
%%%%%%%%%%%%%%%%%%%%%%%%%%%%%%%%%%%%%%%%%%%%%%%%%%%%%%%%%%%%%%%%%%%%%%%%%%%%%%%%%%%%%%%%%
\section{Introduction}
Examination of form factors calculated in different forms 
of relativistic quantum mechanic (RQM) \cite{Dirac:1949cp,Keister:sb}, 
with the same solution of a mass operator, shows tremendous differences 
depending on the approach that is used \cite{Amghar:2002jx, Desplanques:2004sp}. 
On the other hand, fitting the wave function to experiment shows 
that it depends strongly on the approach \cite{He:2004ba,Julia-Diaz:2004gq}. 
At first sight, one cannot therefore decide whether some discrepancy 
is due to the chosen implementation of relativity or to the underlying dynamics. 
It is however believed that results should not depend on the choice 
of the hypersurface underlying some form of relativity 
\cite{Sokolov:1978} 
and that their equivalence requires the introduction of many-body currents.

Among properties expected from Poincar\'e covariance, invariance of form
factors under rotations or boosts can easily be checked 
by applying these transformations to the system under consideration. 
In absence of a similar test, the role of space-time translations, 
which are also part of the Poincar\'e group, is much less known 
beyond the energy-momentum conservation that is, of course, 
assumed to hold. In this contribution, we examine this role 
and show that accounting for the constraints they imply can remove 
large discrepancies between results obtained in different approaches 
with the simplest one-body current.

In the following, we successively consider the space-time translations 
and constraints they imply (sect. 2), the implementation 
of these constraints (sect. 3), the illustration of their role 
for a pion-like system (sect. 4) and, finally, the relation 
to a dispersion-relation approach that could be considered 
as the convergence point of all approaches based on the same solution 
of a mass operator (sect. 5). 
Due to lack of space, details had to be skipped. 
They could be found in ref. \cite{Desplanques:2008fg}.
%%%%%%%%%%%%%%%%%%%%%%%%%%%%%%%%%%%%%%%%%%%%%%%%%%%%%%%%%%%%%%%%%%%%%%%%%%%%%%%%%%%%%%%%%
%%%%%%%%%%%%%%%%%%%%%%%%%%%-2222222222-%%%%%%%%%%%%%%%%%%%%%%%%%%%%%%%%%%%%%%%%%%%%%
%%%%%%%%%%%%%%%%%%%%%%%%%%%%%%%%%%%%%%%%%%%%%%%%%%%%%%%%%%%%%%%%%%%%%%%%%%%%%%%%%%%%%%%%%
\section{Constraints from Poincar\'e space-time translation invariance}  
Under space-time translations, a vector or a scalar current transforms 
as follows:
\begin{eqnarray}
e^{iP \cdot a}\;J^{\nu}(x) \;({\rm or}\;S(x))\;e^{-iP \cdot a}=J^{\nu}(x+a) \;({\rm or}\;S(x+a)),
\label{eq:translat1}
\end{eqnarray}
where $P^{\mu}$ represents the total momentum operator. Considering the matrix
element of this equality between eigenstates of $P^{\mu}$ for $a=-x$, one gets:
\begin{eqnarray}
<i\;| J^{\nu}(x) \;({\rm or}\;S(x))|\;f>=
e^{i\,(P_i-P_f) \cdot x}\;<i\;|J^{\nu}(0) \;({\rm or} \;S(0))|\;f>.
\label{eq:translat2}
\end{eqnarray}
Together with a field carrying momentum $q^{\mu}$, one gets under the assumption of
space-time translation invariance the current momentum-energy conservation,
$(P_f-P_i)^{\mu} = q^{\mu}$.

\begin{figure}[htb]
\mbox{ \hspace*{4cm}\epsfig{ file=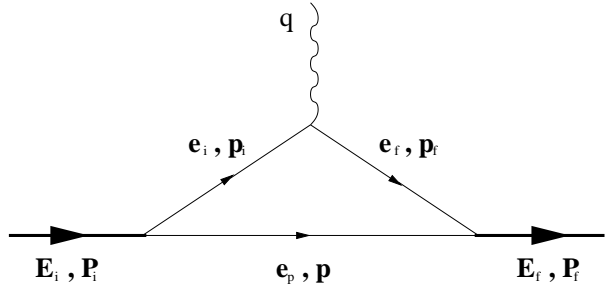, width=8cm}}
\caption{Representation of the interaction with an external probe
\label{fig:probe}}
\end{figure}

For the purpose of calculating form factors, $J^{\nu}(0)\; ({\rm or}\; S(0))$ 
is generally approximated by a one-body current. 
Equation (\ref{eq:translat1}) however implies further
relations involving the commutator of $P^{\mu}$ with the currents
and the derivative of the current with respect to $x$ \cite{Lev:1993}. 
Particularly interesting relations are the following double commutators:
\begin{eqnarray}
\hspace*{1cm}\Big[P_{\mu}\;,\Big[ P^{\mu}\;,\; J^{\nu}(x)\Big]\Big]=
-\partial_{\mu}\,\partial^{\mu}\,J^{\nu}(x),
\;\;\; 
\Big[P_{\mu}\;,\Big[ P^{\mu}\;,\; S(x)\Big]\Big]=
-\partial_{\mu}\,\partial^{\mu}\,S(x) \, . 
\end{eqnarray}
Considering the matrix element of these relations  between eigenstates 
of $P^{\mu}$, and for the case of a one-body current, one should verify 
the relation:
\begin{eqnarray}
<\;|q^2\; J^{\nu}(0) \;({\rm or}\;S(0))|\;>=
<\;|(p_i-p_f)^2\,J^{\nu}(0)\;({\rm or}\;S(0))|\;> \,. 
\end{eqnarray}
It is easily seen that this equation cannot be generally fulfilled, 
as most often $q^2 \neq (p_i-p_f)^2$ in RQM approaches 
(see fig. \ref{fig:probe} for a
graphical representation). This implies
that the assumption of a single-particle current is inconsistent 
with properties from space-time translations and that the current, 
$J^{\nu}(0)\; ({\rm or}\; S(0))$, besides a one-body component,  
should also contain many-body components which, until now, have been ignored.

%
%%%%%%%%%%%%%%%%%%%%%%%%%%%%%%%%%%%%%%%%%%%%%%%%%%%%%%%%%%%%%%%%%%%%%%%%%%%%%%%%%%%%%%%%%
%%%%%%%%%%%%%%%%%%%%%%%%%%%%%-33333333-%%%%%%%%%%%%%%%%%%%%%%%%%%%%%%%%%%%%%%%%%%%%%%
%%%%%%%%%%%%%%%%%%%%%%%%%%%%%%%%%%%%%%%%%%%%%%%%%%%%%%%%%%%%%%%%%%%%%%%%%%%%%%%%%%%%%%%%%
\section{Implementation of the constraints}
It is expected that many-body currents at all orders in the interaction
are required to fulfill constraints from space-time translations. 
As considering these currents explicitly is excluded, we consider them 
indirectly, by modifying wave functions and the current operator. 
The modification is suggested by examining expressions of form factors 
in various approaches, which show that the factor multiplying $Q$ 
could be given by a factor varying from 1 in some cases to $\frac{2e_k}{M}$ 
in other ones \cite{Desplanques:2003nk}. 
The departure of this quantity to 1 represents an interaction term 
and is a signature of the hypersurface underlying the approach. 
To account for the constraints, we thus propose to multiply $Q$
by a factor $\alpha$ and determine this one by requiring 
that the squared momentum transferred to the system, $q^2$, 
be equal to the one for the constituents,  denoted $``(p_i-p_f)"^2$. 
The equation to be solved is typically given by:
\begin{eqnarray}
q^2&=&
``[(P_i\!-\!P_f)^2+2\, (\Delta_i \!-\! \Delta_f)\;  (P_i\!-\!P_f) \cdot \xi
+(\Delta_i \!-\!\Delta_f)^2 \;\xi^2]"
\nonumber \\
&=&\alpha^2q^2-2\,\alpha\; ``(\Delta_i \!-\!\Delta_f)" \;q \cdot \xi
 +``(\Delta_i \!-\!\Delta_f)^2"\;\xi^2 \,,
\end{eqnarray}
where $\Delta$, which represents an interaction effect, also depends 
on $\alpha$. Explicit expressions of $\alpha$ can be found 
in ref. \cite{Desplanques:2008fg} for different forms.  
Expressions for form factors, taking into account the effect of constraints
motivated by space-time translation properties, 
can then be obtained  \cite{Desplanques:2008fg}. 
By expanding these expressions in terms of  $\Delta$, the many-body character 
of corrections at all orders of the interaction could be checked.

A few points deserve to be noticed. Firstly, for the standard front-form 
where $\xi^2=0$, $q\cdot\xi\;({\rm or}\; q^+)=0$, the factor $\alpha$ 
is equal to 1 and, therefore, results for the form factors are unchanged.
Secondly, a solution has been found but we do not exclude that the implementation 
can be done differently, what would be desirable to fulfill the infinite set of
constraints involving the multiple commutators of $P^{\mu}$ with currents.
Thirdly, the choice of the charge current is constrained for some part. 
We used for our purpose a current inspired from results for the simplest Feynman triangle
diagram. This allows one to get Lorentz invariant results for form factors.
%%%%%%%%%%%%%%%%%%%%%%%%%%%%%%%%%%%%%%%%%%%%%%%%%%%%%%%%%%%%%%%%%%%%%%%%%%%%%%%%%%%%%%%%%
%%%%%%%%%%%%%%%%%%%%%%%%%%%-4444444444-%%%%%%%%%%%%%%%%%%%%%%%%%%%%%%%%%%%%%%%%%%
%%%%%%%%%%%%%%%%%%%%%%%%%%%%%%%%%%%%%%%%%%%%%%%%%%%%%%%%%%%%%%%%%%%%%%%%%%%%%%%%%%%%%%%%%
\section{Numerical illustration}
\begin{figure}[htb]
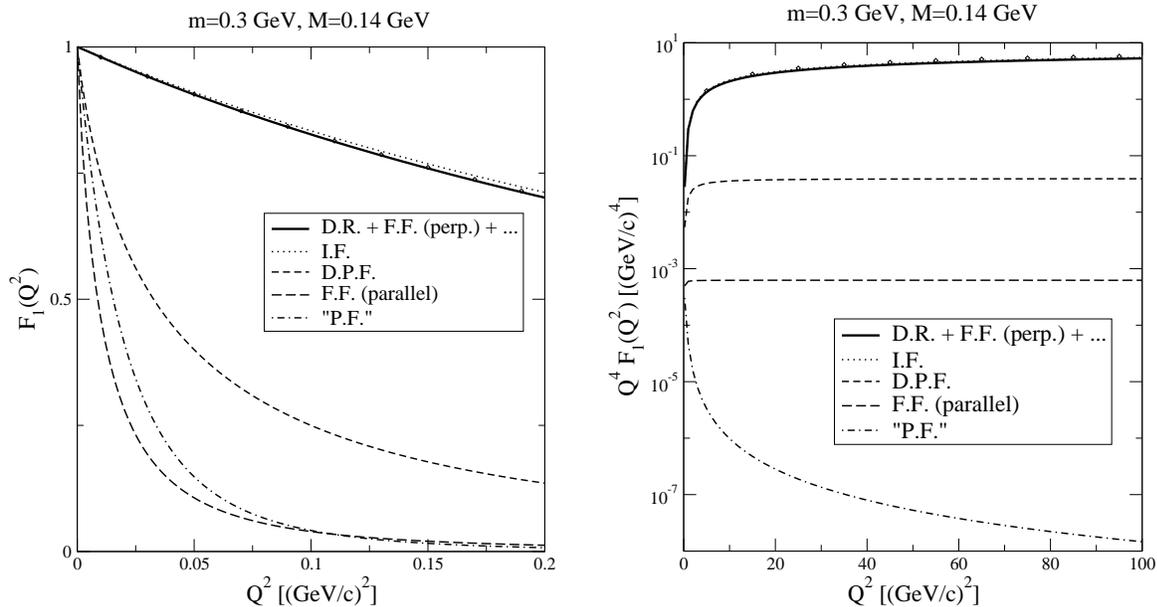

\mbox{ \epsfig{ file=frat1s.eps, width=7.3cm}}
\hspace*{0.4cm}
\mbox{\epsfig{ file=frat1S.eps, width=7.3cm}}
\caption{Charge form factors. See text for the curves. 
 The ``experiment" is represented by diamonds. 
\label{fig:charge}}
\end{figure} 
To illustrate effects of the restoration of properties related to space-time 
translations, we consider a system of scalar particles interacting through 
the exchange of massless particles (Wick-Cutkosky model 
\cite{Wick:1954,Cutkosky:1954}). 
For this system, that may be used as a test case \cite{Karmanov:1991fv}, 
form factors can be calculated exactly. There are two of them 
for the ground state, a charge and a scalar one, 
which provide us with some ``expe\-ri\-ment". The RQM calculations are based 
on the solution of a mass operator where the interaction is chosen 
so that to reproduce approximately the degeneracy pattern 
of the exact spectrum while keeping the high-momentum power law unchanged. 
The strength is fitted to reproduce the energy of the ground state 
used in the calculations. The mass of the system and of the constituents 
are those currently used for the pion, $M=0.14$ GeV and $m=0.3$ GeV.
\begin{figure}[htb]
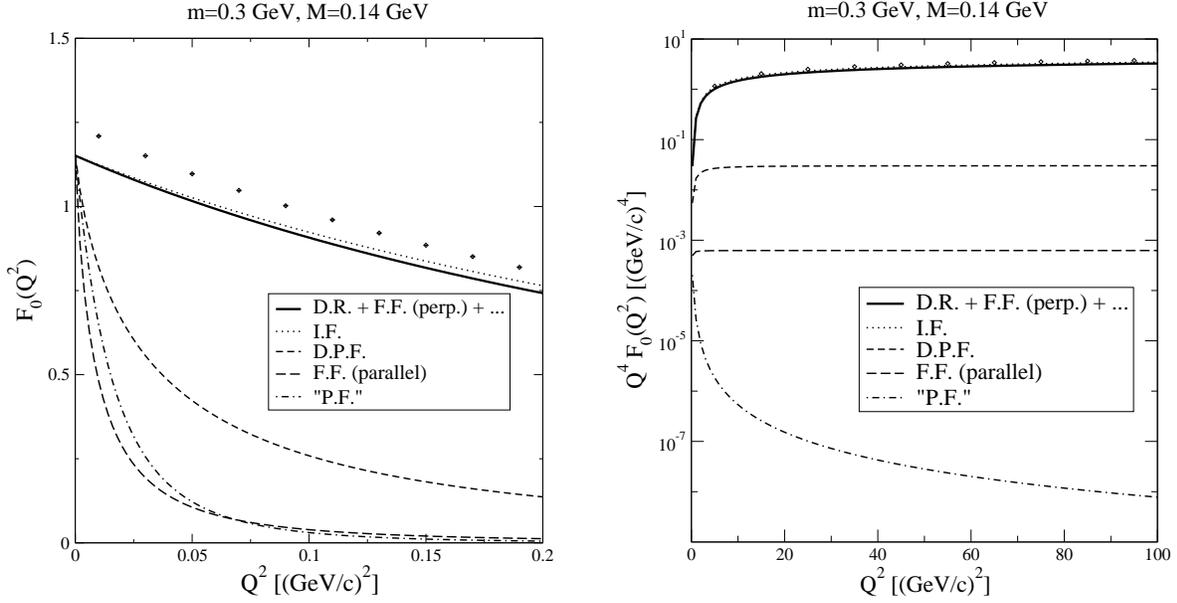

\mbox{ \epsfig{ file=frat0s.eps, width=7.3cm}}
\hspace*{0.4cm}
\mbox{ \epsfig{ file=frat0S.eps, width=7.4cm}}
\caption{Scalar form factors. See text for the curves. 
 The ``experiment" is represented by diamonds.
\label{fig:scalar}}
\end{figure} 

Results are presented in figs. \ref{fig:charge}, \ref{fig:scalar} 
for the charge and scalar form factors. 
In each case, there are two panels describing the low- and high-$Q^2$ behavior, 
respectively sensitive to the square radius and to the asymptotic behavior. 
The different curves correspond to uncorrected Breit-frame form factors 
for the front form ($q^+=0$; F.F. (perp.)), a front form 
with a parallel configuration (F.F. (parallel)), the usual instant form (I.F.) 
\cite{Bakamjian:1953kh}, 
some instant form with the symmetry properties of the point form  (``P.F.")
\cite{Bakamjian:1961}, 
a point form inspired from the Dirac one (D.P.F.) \cite{Desplanques:2004rd}, 
and corrected results 
which, all, coincide with the standard front-form results ($q^+=0$) 
that are unchanged. Anticipating on next section, we also show results 
of a dispersion-relation approach (D.R.), which coincide with the standard 
front-form ones.

Examination of results for the charge form factor (fig. \ref{fig:charge}) 
shows that tremendous discre\-pancies are removed by accounting for constraints 
from space-time translation properties at both low and high $Q^2$. 
The paradox of a charge radius tending to infinity while the mass 
of the system goes to zero (or the interaction is increased) disappears. 
There remain a slight discrepancy with ``experiment". 
It can be ascribed for a part to the description of the mass operator. 
It is not clear whether genuine two-body currents are needed. 
Exa\-mination of results for the scalar form factor (fig. \ref{fig:scalar}) 
shows similar features as far as corrections related 
to space-time translation properties are concerned. 
In contrast to the charge form factor, genuine two-body currents 
are needed here to explain the form factor at low $Q^2$. It is reminded that, 
contrary to the charge form factor, the scalar one at $Q^2=0$ 
is not protected by some conservation law.
%%%%%%%%%%%%%%%%%%%%%%%%%%%%%%%%%%%%%%%%%%%%%%%%%%%%%%%%%%%%%%%%%%%%%%%%%%%%%%%
%%%%%%%%%%%%%%%%%%%%%%%%%%%-555555555-%%%%%%%%%%%%%%%%%%%%%%%%%%%%%%%%%%%%
%%%%%%%%%%%%%%%%%%%%%%%%%%%%%%%%%%%%%%%%%%%%%%%%%%%%%%%%%%%%%%%%%%%%%%%%%%%%%%
\section{Relation to a dispersion-relation approach}
A dispersion-relation has been proposed to calculate form factors 
\cite{Anisovich:1992hz,Krutov:2001gu,Melikhov:2001zv}. The
expressions for the charge and scalar form factors read: 
\begin{eqnarray} 
F_1(Q^2) &\!=\!& \frac{1}{N} \int \!\!\int 
d\bar{s}\,d\Big(\frac{s_i\!-\!s_f}{Q}\Big) \; 
\frac{ \,(2\bar{s}+Q^2)\;\theta\Big(\frac{s_i\;s_f}{D} -m^2\Big) }{
D^{3/2} }\; \phi(s_f) \, \phi(s_i)\,, 
\nonumber \\
F_0(Q^2) &\!=\!& \frac{1}{N} \int \!\! \int   
d\bar{s}\,d\Big(\frac{s_i\!-\!s_f}{Q}\Big) \; 
 \frac{\;\theta\Big(\frac{s_i\;s_f}{D} -m^2\Big)}{2\, D^{1/2} } \;
 \phi(s_f) \, \phi(s_i)  \, , 
\nonumber \\
{\rm where} \;\;\;N &=& \int ds \;\sqrt{\frac{s-4\,m^2}{s}}\;\phi^2(s) \,,
 \hspace*{0.5cm} 
\bar{s}=\frac{s_i\!+\!s_f}{2} \, , \hspace*{0.5cm} 
D=4\bar{s}\!+\!Q^2\!+\!\frac{(s_i\!-\!s_f)^2}{Q^2}\, .
\end{eqnarray}
This approach was presented as an instant-form one \cite{Krutov:2001gu}. 
In absence of reference to a particular direction, it could be thought 
as a  point-form one. Moreover,  one of the authors (B.D.) found that 
the above expressions could be obtained from the standard 
front-form ones ($q^+=0$) by an appropriate change of variables. 
A similar result was obtained independently by Melikhov \cite{Melikhov:2001zv} 
for the pion case.
Examining the approach, it is noticed that it is based on the free 
scattering amplitude in an external field. 
The interaction effects that are here or there in RQM approaches 
are therefore absent. The variables $s_i\!=\!(p_i\!+\!p)^2$, 
$s_f\!=\!(p_f\!+\!p)^2$ and $(p_i-p_f)^2$ 
refer to on-mass shell constituents. 
The functions $\phi(s)=\tilde{\phi}(k^2=\frac{s}{4}-m^2)$ 
can therefore be identified to the solution of a mass operator used 
in RQM approaches. The relation $q^2=(p_i-p_f)^2$, which ensures 
that the square  momentum transferred to the system  be equal 
to the one transferred to the constituents, fulfills the constraints 
expected from space-time translation properties. These three features 
suggest that the RQM results for form factors should converge 
to the dispersion-relation ones, once the above constraints are accounted for.

As the relations of corrected RQM form factors to the dispersion-relation ones 
are not well known,  we give here expressions evidencing the relation 
for the charge form factor. The explicit form of the change of variables 
can be found in ref. \cite{Desplanques:2008fg}. The first case refers to the standard front 
form approach ($q^+=0$), which is of particular interest 
for this light-cone meeting. Using the $k_{\perp}$ and $x$ variables 
currently employed in this domain, the relation reads:

\begin{eqnarray}
F_1(Q^2)&\!=\!&\frac{2}{\pi N}\! \int\! 
\frac{d^2p_{\perp}\;dx}{2x(1\!-\!x)}\;
\tilde{\phi}(\vec{k_f}^2)\;\tilde{\phi}(\vec{k_i}^2) 
\nonumber \\ 
 & \!=\! & \frac{2}{\pi N} \int \! \!\int \!d\bar{s} \; 
d(\frac{s_i\!-\!s_f}{Q}) \frac{1}{2\, \sqrt {D}}\;\phi(s_f) \; \phi(s_i)
%\nonumber \\ && \times
\times\!\int \!\frac{  dx \;\Big((2\,\bar{s}\!+\!Q^2)/D-(x\!-\!d)\Big)}{
\sqrt {\Big  (\frac {s_i\,s_f}{D }\!-\!m^2\Big )f-(x\!-\!d)^2}   }
\nonumber \\ 
& = &\frac{1}{N} \int  \! \!\int d\bar{s} \; 
d(\frac{s_i\!-\!s_f}{Q})  \;  
\frac{ \;(2\,\bar{s}+Q^2) \,\theta(\cdots)}{
D^{3/2} } \;\phi(s_f) \; \phi(s_i) \,.
\end{eqnarray}
The second case refers to an arbitrary front orientation $\lambda^{\mu}$ 
with finite $\lambda^2$. The relation reads:
\begin{eqnarray}
F_1(Q^2)=\frac{16\pi^2}{N(2\pi)^3}\!
\int \frac{d\vec{p}}{e_p}\,\frac{``(2p\!+\!p_i\!+\!p_f)"\!\cdot\! \lambda 
}{2``(p_i\!+\!p_f)" \!\cdot\! \lambda}\;
``\tilde{\phi}(\vec{k_f}^2)\;\tilde{\phi}(\vec{k_i}^2)" \hspace*{5cm}
\nonumber \\
=\!\frac{2}{\pi N}\!
\int \! \!\int \! d\bar{s} \; d(\frac{s_i\!-\!s_f}{Q})\;
\frac{\theta(\cdots) }{4\,D^{3/2}}\; \phi(s_f) \; \phi(s_i)
\!\times \!\sum \!\int \!\frac { d(p \!\cdot\! \hat{\lambda})
\Big ((2\bar{s}\!+\! Q^2)-(p\!\cdot\! \hat{\lambda}\!-\!d)g\Big)
}{\sqrt{\Big( \frac{s_is_f}{D} \!-\!m^2\Big)f \!-\! 
(p\!\cdot\! \hat{\lambda}\!-\!d)^2}}\, .
\end{eqnarray}
In this case, it is noticed that the dependence on $\lambda^{\mu}$ 
and $\bar{P}^{\mu}=(P_i+P_f)^{\mu}/2$, 
which characterizes quantities $d$, $f$ and $g$  \cite{Desplanques:2008fg}, 
disappears in the integration over $p \cdot \hat{\lambda}$, 
allowing one to get Lorentz-invariant results.
In both cases, the integral is reduced from a three- to a two-dimensional one.
%%%%%%%%%%%%%%%%%%%%%%%%%%%%%%%%%%%%%%%%%%%%%%%%%%%%%%%%%%%%%%%%%%%%%%%%%%%%%%%%%
%%%%%%%%%%%%%%%%%%%%%%%-66666666666-%%%%%%%%%%%%%%%%%%%%%%%%%%%%%%%%%%%%
%%%%%%%%%%%%%%%%%%%%%%%%%%%%%%%%%%%%%%%%%%%%%%%%%%%%%%%%%%%%%%%%%%%%%%%%%%%%%%%%%
\section{Conclusion}
In this contribution, we mainly considered properties related to Poincar\'e 
space-time translations in RQM approaches for the calculation of form factors. 
The current practice is to use these properties to factorize the $x$ dependence 
of the current, allowing one to get the usual energy-momentum  
conservation relation, and to assume that the current at $x=0$ 
is a one-body current. At first sight, there is no relation 
of the 4-momentum transferred to the system and to the constituents 
as in a field-theory approach. Considering further relations implied 
by the transformation of the current under space-time translations, 
we showed that they could not be satisfied with a current reduced 
to a one-body component, pointing to the necessary presence 
of many-body components. These ones allow us to account in  RQM approaches 
for the equality of the momentum transferred to the whole system 
and to the constituents,  which characterizes a field-theory approach.
We described a method to  implement the above many-body contributions 
in the case of a scalar-particle model. When this is done, it is found 
that discrepancies between diffe\-rent RQM approaches for calculating 
form factors can be removed, showing that the role of space-time 
translations extends beyond the standard energy-momentum conservation. 
It is also found that these results could coincide with those 
of a dispersion-relation approach. Altogether, all aspects of the Poincar\'e 
group (rotations, boosts and space-time translations) are essential 
in getting reliable estimates of form factors as far as the implementation of
relativity is concerned. Possible discrepancies with experiment can thus be more
likely ascribed to the underlying dynamics.

Present results can be extended without much difficulty to elastic as well as
inelastic form factors of 0-spin systems consisting of 1/2-spin constituents 
with unequal masses. Extension to non-zero spin systems could require more
elaboration.

One of the authors (B.D.) is very grateful to V.A. Karmanov and W. Polyzou for
an initiation to implementation of relativity in field theory and relativistic
quantum mechanic respectively. It is likely that the complementary knowledge of
these two approaches was essential in obtaining results presented here.
This work is partly supported by the National Sciences Foundations of China
under grant No. 10775148.

\end{document}